\documentclass[global]{svjour}
\usepackage{amssymb}
\usepackage{amsfonts}
\usepackage{amsmath}

\input{tcilatex}

\begin{document}

\journalname{Int J Theor Phys}
\title{Hawking Radiation of Linear Dilaton Black Holes in Various Theories}
\subtitle{}
\author{H.Pasaoglu\inst{1} \and I.Sakalli\inst{2}}
\institute{Department of Physics, Eastern Mediterranean University, Gazimagosa, North
Cyprus,\ Mersin 10, Turkey 
\email{hale.pasaoglu@emu.edu.tr}%
\and%
\email{izzet.sakalli@emu.edu.tr}%
}
\headnote{Head note that is usually deleted}
\offprints{Offprints Assistant}
\mail{Address for offprint requests}
\maketitle

\begin{abstract}
Using the Damour-Ruffini-Sannan, the Parikh-Wilczek and the thin film
brick-wall models, we investigate the Hawking radiation\ of uncharged
massive particles from 4-dimensional linear dilaton black holes, which are
the solutions to Einstein-Maxwell-Dilaton, Einstein-Yang-Mills-Dilaton and
Einstein-Yang-Mills-Born-Infeld-Dilaton theories. Our results show that the
tunneling rate is related to the change of Bekenstein-Hawking entropy.
Contrary to the many studies in the literature, here the emission spectrum
is precisely thermal. This implies that the derived emission spectrum is not
consistent with the unitarity of the quantum theory, which would possibly
lead to the information loss.
\end{abstract}

\keywords{Entropy, Linear dilaton black holes, Tunneling effect, Thin film
brick-wall model}

\section{Introduction}

Obeying the laws of black hole mechanics \cite{Bardeen}, Hawking \cite%
{Nature,Commun} proved that a stationary black hole can emit particles from
its event horizon with a temperature proportional to the surface gravity.
According to this idea, the vacuum fluctuations near the horizon would
produce a virtual particle pair, similar to electron-positron pair creation
in a constant electric field. When a virtual particle pair is created just
inside or outside the horizon, the sign of its energy changes as it crosses
the horizon. So after one member of the pair has tunneled to the opposite
side, the pair can materialize with zero total energy. This discovery also
announced the relation between the triple subjects -- the quantum mechanics,
thermodynamics and the gravitation. After this pioneering study of Hawking,
many methods have been proposed to calculate the Hawking radiation for the
last three decades.

One of the commonly used methods is known as Damour-Ruffini-Sannan (DRS) 
\cite{Damour,Sannan} method. This method is applicable to any Hawking
temperature problem in which the asymptotic behaviors of the wave equation
near the event horizon are known.

In 2000, Parikh and Wilczek \cite{ParikhWilczek} proposed a method based on
null geodesics in order to clarify more the Hawking radiation via tunneling
across the event horizon. Namely, they treated the Hawking radiation as a
tunneling process, and used the WKB approximation to determine the
correction spectrum for the black hole's Hawking radiation. In their study,
it is supposed that the barrier depends on the tunneling particle itself.
The crucial point of this method is not to violate the energy conservation
during the process of particle emission and to pass to an appropriate
coordinate system at horizon. In general, the tunneling process is not
precisely a thermal effect and it explains the modification of the black
hole radiation spectrum in which it leads to the unitarity in the quantum
theory \cite{INTParikh,GRGParikh,ArxivParikh}.

Another possible method to study the statistical origin of the black hole
entropy is the brick-wall model initially proposed by t'Hooft \cite{Hooft}.
The brick-wall model identifies the black hole entropy by the entropy of a
thermal gas of quantum field excitations outside the event horizon. Since
then, this method has been satisfactorily applied to many black hole
geometries (see for instance \cite{Liu}, and the references therein).
Although t' Hooft made significant contribution to clarify the understanding
and calculating the entopy of the black holes, there were some drawbacks in
his model. Those drawbacks are overcome by the improved form of the original
brick-wall model, which is called as thin film brick-wall model \cite%
{PRDLiZhao}. The thin film brick-wall model gives us acceptable and net
physical meaning of the entropy calculation. In summary, since the entropy
calculated by the thin film brick-wall model is just from a small region
(thin film) near the horizon, this improved version of the brick-wall model
represents explicitly the correlation between the horizon and the entropy.
In this study, we obtain the ultraviolet cut-off distance as 90$\beta ,$
where $\beta $ is the Boltzmann factor.

Hawking described the black hole radiation as tunneling triggered by the
vacuum fluctuations near the horizon. His discovery, which treats the black
hole radiation as being pure thermal gave also rise to a new paradox in the
black hole physics -- the information loss paradox. Although, Parikh and
Wilczek's tunneling process \cite{ParikhWilczek} is a way to overcome the
information loss paradox in the Hawking radiation, the information might not
be conserved in some black hole geometries. For instance, if only the
tunneling process of the outer horizon of the Reissner-Nordstr\"{o}m black
hole is considered \cite{QJiang,Ren}, it can be shown that the information
loss is possible. The similar violation in the conservation of information
happens in the $4$-dimensional linear dilaton black holes (LDBHs) in various
theories, and we will explain its reason by using the differences in
entropies of the black holes before and after the emission.

The paper is organized as follows: In section 2, a brief overview of the $4$%
-dimensional LDBHs in Einstein-Maxwell-Dilaton (EMD),
Einstein-Yang-Mills-Dilaton (EYMD) and
Einstein-Yang-Mills-Born-Infeld-Dilaton (EYMBID) theories, which they have
been recently employed in \cite{Sakalli} \ for calculating the Hawking
radiation via the method of semi-classical radiation spectrum is given.
Next, we apply the DRS method to find the temperature of the LDBHs and the
tunneling rate of the chargeless particles crossing the event horizon.
Section 3 is devoted to the calculation of the entropy of the horizon by
using all those methods mentioned above. As it is expected, they all
conclude with the same result. Finally, we draw our conclusions and
discussions.

\bigskip Throughout the paper, the units $G=c=$%
h{\hskip-.2em}\llap{\protect\rule[1.1ex]{.325em}{.1ex}}{\hskip.2em}%
$=k_{B}$=1 are used.

\section{LDBHs, Calculation of Their Temperature and Tunneling Rate}

The line-element of\ $N$-dimensional ($N\geq 4$) LDBHs, which are static
spherically symmetric solutions in various theories (EMD, EYMD and EYMBID)
have been recently summarized by \cite{Sakalli}. However, throughout this
paper we restrict ourselves to the $4$-dimensional LDBHs and follow the
notations of \cite{Sakalli}.

Consider a general class of static, spherically symmetric spacetime for the
LDBHs as

\begin{equation}
ds^{2}=-fdt^{2}+\frac{dr^{2}}{f}+A^{2}rd\Omega ^{2},
\end{equation}

where $d\Omega ^{2}=d\theta ^{2}+\sin ^{2}\theta d\phi ^{2}.$ Here, the
metric function $f$ is given by \cite{Sakalli}

\begin{equation}
f=\Sigma r(1-\frac{r_{+}}{r}),
\end{equation}

where $r_{+}$ is the radius of the event horizon. The coefficients $\Sigma $
and $A$ in the metric (1) take different values according to the concerned
theory.

\ Since the present form of the metric represent asymptotically non-flat
solutions, one should consider the quasi-local mass definition $M$ of the
metric (1). In \cite{Sakalli}, the relationship between the horizon $r_{+}$\
and the mass $M$ is explicitly given as%
\begin{equation}
r_{+}=\frac{4M}{\Sigma A^{2}},
\end{equation}

In the EMD theory \cite{Sakalli,Chan,Clement}, the coefficients $\Sigma $
and $A$ are found as 
\begin{equation}
\Sigma \rightarrow \Sigma _{EMD}=\frac{1}{\gamma ^{2}}\text{ \ and \ \ \ }%
A\rightarrow A_{EMD}=\gamma \text{,}
\end{equation}

where $\gamma $ is a constant related to the electric charge of the black
hole. Meanwhile, one can match the metric (1) to the LDBH's metric of Cl\'{e}%
ment \textit{et.al.} \cite{Clement} by setting $\gamma \equiv r_{0}$. Next,
if one considers the EYMD and EYMBID theories \cite{Mazhari1,Mazhari2}, the
coefficients in the metric (1) become

\begin{equation}
\Sigma \rightarrow \Sigma _{EYMD}=\frac{1}{2Q^{2}}\text{ \ and \ \ \ }%
A\rightarrow A_{EYMD}=\sqrt{2}Q,
\end{equation}

and

\begin{equation}
\Sigma \rightarrow \Sigma _{EYMBID}=\frac{1}{Q_{C}^{2}}\left[ 1-\sqrt{1-%
\frac{Q_{C}^{2}}{Q^{2}}}\right] \text{ \ and \ }A\rightarrow A_{EYMBID}=%
\sqrt{2}Q\left( 1-\frac{Q_{C}^{2}}{Q^{2}}\right) ^{\frac{1}{4}},
\end{equation}

where $Q$ and $Q_{C}$ are YM charge and the critical value of YM charge,
respectively. The existence of the metric (1) in EYMBID theory depends
strictly on the condition \cite{Mazhari2}

\begin{equation}
Q^{2}>Q_{C}^{2}=\frac{1}{4\tilde{\beta}^{2}},
\end{equation}

where $\tilde{\beta}$ is the Born-Infeld parameter.\ Meanwhile, it is not
necessary to say that values of $\Sigma $\ in equations\ (4), (5) and (6)
are always positive.

By using the definition of the surface gravity \cite{Wald}, we get

\begin{equation}
\kappa =\lim_{r\rightarrow r_{+}}\frac{f^{\prime }(r)}{2}=\frac{\Sigma }{2}.
\end{equation}

Since the surface gravity (8) is positive, one can deduce that it is
directed towards the singularity. As a consequence, it is attractive and the
matter can only fall into the black hole. This horizon is a future horizon
to an observer, who is located outside of it.

In curved spacetime, a massive test scalar field $\Phi $ with mass $\mu $
obeys the covariant Klein-Gordon (KG) equation, which is given by

\begin{equation}
\frac{1}{\sqrt{-\det g}}\partial _{\mu }\left( \sqrt{-\det g}g^{\mu \nu
}\partial _{\nu }\Phi \right) -\mu ^{2}\Phi =0,
\end{equation}%
\newline

The massive scalar wave equation $\Phi $ in metric (1) can be separated as $%
\Phi =Y(\theta ,\varphi )\psi (t,r)$ in which the radical KG equation (9)
satisfies the following equation:

\begin{equation}
\frac{\partial ^{2}\psi }{\partial t^{2}}+f(\frac{f}{r}+\Sigma )\frac{%
\partial \psi }{\partial r}+f^{2}\frac{\partial ^{2}\psi }{\partial r^{2}}%
-f(\mu ^{2}-\frac{l(l+1)}{r})\psi =0,
\end{equation}

where $l$ is the angular quantum number. In order to change equation (10)
into a standard wave equation at the horizon, we introduce the tortoise
coordinate transformation, which is obtained from

\begin{equation}
dr_{\ast }=\frac{dr}{f},
\end{equation}

After making the straightforward calculation, we find an appropriate $%
r_{\ast }$\ as

\begin{equation}
r_{\ast }=\frac{1}{2\kappa }\ln (r-r_{+}),
\end{equation}

Thus, one can transform the radical equation (10) into the following form

\begin{equation}
\frac{\partial ^{2}\psi }{\partial t^{2}}-\frac{f}{r}\frac{\partial \psi }{%
\partial r_{\ast }}-\Sigma \frac{\partial \psi }{\partial r_{\ast }}+\Sigma 
\frac{\partial \psi }{\partial r_{\ast }}-\frac{\partial ^{2}\psi }{\partial
r_{\ast }^{2}}+f[\mu ^{2}-\frac{l(l+1)}{r}]\psi =0,
\end{equation}

While $r\rightarrow r_{+}$ in which $f\rightarrow 0$, the transformed
radical equation (13) can be reduced to the following standard form of the
wave equation as

\begin{equation}
\frac{\partial ^{2}\psi }{\partial t^{2}}-\frac{\partial ^{2}\psi }{\partial
r_{\ast }^{2}}=0,
\end{equation}

This form of the wave equation reveals that there are propagating waves near
the horizon. The solutions of equation (14), which give us the ingoing and
outgoing waves at the black hole horizon surface $r_{+}$ are

\begin{equation}
\psi _{out}=\exp (-i\omega t+i\omega r_{\ast }),
\end{equation}

\begin{equation}
\psi _{in}=\exp (-i\omega t-i\omega r_{\ast }),
\end{equation}

When we introduce the ingoing Eddington-Finkelstein coordinate, $v=t+r_{\ast
}$, the line-element (1) of the LDBHs becomes

\begin{equation}
ds^{2}=-fdv^{2}+2dvdr+A^{2}rd\Omega ^{2},
\end{equation}

The present form of the metric does not attribute a singularity to the
horizon, so that the ingoing wave equation behaves regularly at the horizon.
This yields the solutions of ingoing and outgoing waves at the horizon $%
r_{+} $ as follows

\begin{equation}
\psi _{out}=e^{-i\omega v}e^{2i\omega r_{\ast }},
\end{equation}

\begin{equation}
\psi _{in}=e^{-i\omega v},
\end{equation}

Now, we consider only the outgoing waves. Namely,

\begin{equation}
\psi _{out}(r>r_{+})=e^{-i\omega v}(r-r_{+})^{\frac{^{i\omega }}{\kappa }},
\end{equation}

which has a singularity at the horizon $r_{+}$. Therefore, equation (20) can
only describe the outgoing particles outside the horizon and strictly cannot
describe the particles, which are inside the horizon. In other words, the
description of the particles' behavior inside horizon has to be made as
well. To this end, the outgoing wave $\psi _{out}$ should be analytically
extended from outside to the interior of the black hole by the lower half
complex $r$-plane

\begin{equation}
(r-r_{+})\rightarrow \left\vert r-r_{+}\right\vert e^{-i\pi
}=(r_{+}-r)e^{-i\pi },
\end{equation}

We can derive the solution of outgoing wave inside the horizon as follows

\begin{equation}
\psi _{out}(r<r_{+})=\psi _{out}^{^{\prime }}\left( r<r_{+}\right) e^{\frac{%
^{\omega \pi }}{\kappa }},
\end{equation}

where

\begin{equation}
\psi _{out}^{^{\prime }}\left( r<r_{+}\right) =e^{-i\omega v}(r_{+}-r)^{%
\frac{^{i\omega }}{\kappa }},
\end{equation}%
\bigskip

According to the Damour-Ruffini-Sannan (DRS) \cite{Damour,Sannan} method, it
is possible to calculate the emission rate. The total outgoing wave function
can be written in a uniform form

\begin{equation}
\psi =N_{\omega }[\Theta (r-r_{+})\psi _{out}\left( r>r_{+}\right) +e^{\frac{%
\omega \pi }{\kappa }}\Theta (r_{+}-r)\psi _{out}^{^{\prime }}\left(
r<r_{+}\right) ],
\end{equation}

where $\Theta $ is the Heaviside step function and $N_{\omega }$ represents
the normalization factor. From the normalization condition

\begin{equation}
\left( \psi ,\psi \right) =\pm 1,
\end{equation}

we can obtain the resulting radiation spectrum of scalar particles

\begin{equation}
N_{\omega }^{2}=\frac{\Gamma }{1-\Gamma }=\frac{1}{e^{\frac{\omega }{T}}-1},
\end{equation}

and read the temperature of the horizon as

\begin{equation}
T=\frac{\kappa }{2\pi },
\end{equation}

In equation (26) $\Gamma $ symbolizes the emission or tunneling rate, which
is found by the following ratio

\begin{equation}
\Gamma =\left\vert \frac{\psi _{out}\left( r>r_{+}\right) }{\psi
_{out}\left( r<r_{+}\right) }\right\vert ^{2}=e^{\frac{-^{2\pi \omega }}{%
\kappa }}.
\end{equation}

One can remark for this section that the resulting temperature (27) obtained
from the DRS method is in agreement with the statistical Hawking temperature 
\cite{Wald} computed as usual by dividing the surface gravity by $2\pi $.

\section{Entropy of the Horizon}

In this section, we shall use three different methods in order to show that
they all lead to the same entropy result. We first employ the DRS method,
which is worked in detail and obtained remarkable results in the previous
section. The second method will be the Parikh-Wilczek method \cite%
{ParikhWilczek} describing the Hawking radiation as a tunneling process.
Last method that is also going to be used in the calculation of the entropy
is the thin film brick-wall model \cite{PRDLiZhao}.

In the DRS method, the emission rate of outgoing particles is found as in
equation (28). Accordingly, the probability of emission can be modified into 
\cite{QJiang,Jiang} (and references therein)

\begin{equation}
\Gamma =e^{^{-2\pi \int_{0}^{\omega }\frac{d\omega ^{\prime }}{\kappa }%
}}=e^{-\int_{0}^{\omega }\frac{d\omega ^{\prime }}{T}}=e^{\Delta S_{BH}},
\end{equation}

where $\Delta S_{BH}$ is the difference of Bekenstein-Hawking entropies of
the LDBHs before and after the emission of the particle.

On the other hand, the novel study on the tunneling effect is designated by
Parikh-Wilczek method \cite{ParikhWilczek}, which proposes an approach for
calculating the tunneling rate at which particles tunnel across the event
horizon. They treated Hawking radiation as a tunneling process, and used the
WKB method \cite{ArxivParikh}. In classical limit, we can also find the
tunneling rate by applying WKB approximation. This relates the tunneling
amplitude to the imaginary part of the particle action at stationary phase
and the Boltzmann factor for emission at the Hawking temperature.

In the WKB approximation, the imaginary part of the amplitude for outgoing
positive energy particle which crosses the horizon outwards from initial
radius of the horizon $r_{in}$\ to the final radius of the horizon $r_{out}$%
\ could be expressed by

\begin{equation}
\func{Im}I=\func{Im}\int_{r_{in}}^{r_{out}}p_{r}dr=\func{Im}%
\int_{r_{in}}^{r_{out}}\int_{0}^{p_{r}}dp_{r}^{\prime }dr,
\end{equation}

By using the standard quantum mechanics, the tunneling rate $\Gamma $ is
given in the WKB approximation as \cite{KrausWilczek1,KrausWilczek2},

\begin{equation}
\Gamma \sim \exp (-2\func{Im}I),
\end{equation}

Here we can consider the particle with energy $\omega $ as a shell of energy
and fix the total mass $M$ (quasi-local mass) and allow the hole mass to
fluctuate. Then the Hamilton's equation of motion can be used to write $%
dp_{r}=\frac{dH}{\dot{r}},$ and it can be noted that the horizon moves
inwards from $M$ to $M-\omega $\ while a particle emits. Introducing $%
H=M-\omega $\ \ and inserting the value of the\ $\dot{r}\equiv \frac{dr}{dv}=%
\frac{f}{2}$ obtained from the null geodesic equation into (30), we obtain

\begin{equation}
\func{Im}\int_{r_{in}}^{r_{out}}\int_{0}^{p_{r}}dp_{r}^{\prime }dr=\func{Im}%
\int_{r_{in}}^{r_{out}}\int_{M}^{M-\omega }\frac{dr}{\dot{r}}dH=\func{Im}%
\int_{0}^{\omega }\int_{r_{in}}^{r_{out}}\frac{2dr}{\Sigma (r-r_{+})}\left(
-d\omega ^{\prime }\right) ,
\end{equation}

The $r$-integral can be done by deforming the contour. The deformation of
the integral is based on an assumption that the contour semicircles the
residue in a clockwise fashion. In this way, one can obtain

\begin{equation}
\func{Im}I=2\pi \int_{0}^{\omega }\frac{d\omega ^{\prime }}{\Sigma },
\end{equation}

So, the tunneling rate (31) is

\begin{equation}
\Gamma \sim \exp (-2\func{Im}I)=\exp \left( -4\pi \int_{0}^{\omega }\frac{%
d\omega ^{\prime }}{\Sigma }\right) =\exp \left( \Delta S_{BH}\right) ,
\end{equation}

Our result (34) is consistent with the results of the other works \cite%
{ParikhWilczek,Kerner,Zhang,Chen,Li}.

Now, we come to the stage to apply the thin film brick-wall model \cite%
{PRDLiZhao}, which was based on the brick wall model proposed firstly by
t'Hooft \cite{Hooft}. According to this model, the considered field outside
the horizon is assumed to be non-zero only in a thin film, which exists in a
small region bordered by $r_{+}+\varepsilon $ and $r_{+}+\varepsilon +\delta 
$. Here, $\varepsilon $\ is the ultraviolet cut-off distance and $\delta $
is the thickness of the thin film. In summary, both $\varepsilon $ and $%
\delta $ are positive infinitesimal parameters. This model treats the
entropy as being associated with the field in the considered small region in
which the local thermal equilibrium and the statistical laws are valid \cite%
{Tian}. That is why one can work out the entropy of the horizon by using
this model.

If one redefines the massive test scalar field $\Phi $ as being $\Phi
=e^{-i\omega t}\psi (r)Y_{lm}(\theta ,\phi )$ in the KG equation (9) and
considers its radial part only, the wave vector is found with the help of
WKB approximation as

\begin{equation}
k^{2}=\frac{1}{\Sigma r(1-\frac{r_{+}}{r})}[\frac{\omega ^{2}}{\Sigma r(1-%
\frac{r_{+}}{r})}-(\mu ^{2}+\frac{l(l+1)}{A^{2}r})],
\end{equation}

Using the quantum statistical mechanics, we calculate the free energy from

\begin{equation}
F=\frac{-1}{\pi }\int_{0}^{\infty }d\omega \int_{r}dr\int_{l}(2l+1)\frac{k}{%
e^{\beta \omega }-1}dl,
\end{equation}

While integrating equation (36) with respect to $l,$ one should consider the
upper limit of integration such that $k^{2}$ remains positive, and the lower
limit becomes zero. Briefly, we get

\begin{equation}
F\cong \frac{-2A^{2}}{3\pi \Sigma ^{2}}\int_{0}^{\infty }\frac{d\omega }{%
e^{\beta \omega }-1}\int_{r}\frac{r}{(r-r_{+})^{2}}[\omega ^{2}-\mu
^{2}\Sigma (r-r_{+})]^{\frac{3}{2}}dr,
\end{equation}

where $\beta $\ denotes the inverse of the temperature. In equation (37),
the integration with respect to $r$ is quite difficult. On the other hand,
the thin film brick-wall model imposes us to take only the free energy of a
thin layer near horizon of a black hole, and the integration with respect to 
$r$ must be limited in the region $r_{+}+\varepsilon \leq r\leq
r_{+}+\varepsilon +\delta .$ The natural result of this choice sets the
coefficient of $\mu ^{2}$ to zero, and whence the integration of equation
(37) with respect to $\omega $ becomes very simple such that it can be
easily found as $\pi ^{4}/15\beta ^{4}$. Finally, the equation (37) reduces
to

\begin{equation}
F\cong \frac{-2\pi ^{3}A^{2}}{45\beta ^{4}\Sigma ^{2}}\int_{r_{+}+%
\varepsilon }^{r_{+}+\varepsilon +\delta }\frac{r}{(r-r_{+})^{2}}dr,
\end{equation}

\begin{equation}
\cong \frac{-2\pi ^{3}A^{2}r_{+}}{45\beta ^{4}\Sigma ^{2}}%
\int_{r_{+}+\varepsilon }^{r_{+}+\varepsilon +\delta }\frac{dr}{(r-r_{+})^{2}%
},
\end{equation}

\begin{equation}
F\cong \frac{-2\pi ^{3}A^{2}r_{+}}{45\beta ^{4}\Sigma ^{2}}\frac{\delta }{%
\varepsilon (\delta +\varepsilon )},
\end{equation}

and we can get the entropy

\begin{equation}
S_{BH}=\beta ^{2}\frac{\partial F}{\partial \beta }=[\frac{8\pi
^{3}A^{2}r_{+}}{45\beta ^{3}\Sigma ^{2}}]\frac{\delta }{\varepsilon (\delta
+\varepsilon )},
\end{equation}

Since the beta is the inverse of the temperature

\begin{equation}
\beta =\frac{1}{T}=\frac{4\pi }{\Sigma },
\end{equation}

and if we select an appropriate cut-off distance $\varepsilon $ and
thickness of thin film $\delta $\ to satisfy

\begin{equation}
\frac{\delta }{\varepsilon (\delta +\varepsilon )}=90\beta ,
\end{equation}%
the total entropy of the horizon becomes

\begin{equation}
S_{BH}=\frac{1}{4}A_{h},
\end{equation}

where $A_{h}$ is the area of the the black hole horizon, i.e. $A_{h}=4\pi
A^{2}r_{+}.$ The derivative of the entropy (44) with respect to $M$ is

\begin{equation}
\frac{\partial S_{BH}}{\partial M}=\pi A^{2}\frac{\partial r_{+}}{\partial M}%
=\frac{4\pi }{\Sigma },
\end{equation}

Getting the integral of $M$, equation (45) becomes to

\begin{equation}
\Delta S_{BH}=\int_{M}^{M-\omega }\frac{\partial S_{BH}}{\partial M^{\prime }%
}dM^{\prime }=4\pi \int_{M}^{M-\omega }\frac{dM^{\prime }}{\Sigma },
\end{equation}

After substituting $M^{\prime }=M-\omega ^{\prime }$\ into the above
equation, we obtain

\begin{equation}
\Delta S_{BH}=-\int_{0}^{\omega }\frac{\partial S_{BH}}{\partial M^{\prime }}%
d\omega ^{\prime }=-4\pi \int_{0}^{\omega }\frac{d\omega ^{\prime }}{\Sigma }%
,
\end{equation}%
\ 

One can easily see that equation (47) is nothing but the results obtained
both from Parikh-Wilczek method (34) and the DRS method (29).

On the other hand, for the LDBHs the change of the entropy before and after
the radiation is

\begin{equation}
\Delta S_{BH}=S(M-\omega )-S(M)=-\frac{2\pi \omega }{\kappa }.
\end{equation}

Since equation (48) contains only $\omega $, we deduce that the spectrum is
precisely thermal. In other words, the thermal spectrum does not suggest the
underlying unitary theory, and whence we can understand that the
conservation of information is violated.

\section{Discussion and Conclusion}

In this paper, we have effectively utilized three different methods (the DRS
model, the Parikh-Wilczek model and the thin film brick wall model) to
investigate the Hawking radiation for massive 4-dimensional LDBHs in the
EMD, EYMD and EYMBID theories. By considering the DRS method, the tunneling
probability for an outgoing positive energy particle or simply the tunneling
rate is neatly found. Later on, it is shown that the tunneling rate found
from the DRS method can be expressed in terms of the difference of
Bekenstein-Hawking entropies $\Delta S_{BH}$ of the black holes. Beside
this, the other two methods i.e. the Parikh-Wilczek method and the thin film
brick-wall model attribute also the same $\Delta S_{BH}$ result. In the thin
film brick-wall model, the cut-off factor is found to be 90$\beta $, which
is exactly same as in the calculation of the entropy for the Schwarzschild
black hole \cite{Liancheng}.

On the other hand, the obtained$\ \Delta S_{BH}$ result shows us that the
emission spectrum is nothing but a pure thermal spectrum. This result is not
consistent with the unitarity principle of quantum mechanics. It also
implies the violation of the conservation of information in the LDBHs.

Finally, further application of the Hawking radiation of the charged massive
particles via different methods to the case of LDBHs in higher dimensions 
\cite{Sakalli} may reveal more information compared to the present case.
This will be our next problem in the near future.

\end{document}